# Applying Conceptual Blending to Model Coordinated Use of Multiple Ontological Metaphors

Benjamin W. Dreyfus[1], Ayush Gupta[1], and Edward F. Redish

*Department of Physics, University of Maryland, College Park, MD, USA*

Energy is an abstract science concept, so the ways that we think and talk about energy rely heavily on ontological metaphors: metaphors for what kind of thing energy is. Two commonly used ontological metaphors for energy are *energy as a substance* and *energy as a vertical location*. Our previous work has demonstrated that students and experts can productively use both the substance and location ontologies for energy. In this paper, we use Fauconnier and Turner's conceptual blending framework to demonstrate that experts and novices can successfully blend the substance and location ontologies into a coherent mental model in order to reason about energy. Our data come from classroom recordings of a physics professor teaching a physics course for the life sciences, and from an interview with an undergraduate student in that course. We analyze these data using predicate analysis and gesture analysis, looking at verbal utterances, gestures, and the interaction between them. This analysis yields evidence that the speakers are blending the substance and location ontologies into a single blended mental space.

## Introduction

Energy is a central concept across the sciences in physics, biology, and chemistry. Understanding how students think about energy is essential to science education within each discipline and to making connections across disciplines. In our previous work (Dreyfus, Sawtelle, Turpen, Gouvea, & Redish, 2014b; Dreyfus et al., 2014c), we examined the conceptual demands placed by an interdisciplinary approach to energy in introductory physics, specifically when trying to understand the concept of negative energy. Negative energy becomes an important tool for understanding chemical bond energy (an important concept in introductory biology and chemistry) in terms of potential energy (an important concept in introductory physics). In many introductory physics courses, chemical energy is not an explicit topic of instruction, but becomes so when the course content is reformed with an emphasis on seeking interdisciplinary coherence between physics, biology, and chemistry (Redish et al., 2014; Dreyfus et al., 2014a). This poses instructional challenges.

Recently, many researchers who focus on developing learners' conceptual understanding of energy have emphasized the advantages of using a substance metaphor for energy. Conceptualizing *energy as a substance* is instructionally helpful because it tacitly brings along many useful properties in terms of accounting for the transfer and conservation of energy. However, it is difficult to extend the *energy as substance* metaphor to negative energy because we typically don't conceptualize substances as "negative" or less than nothing. However, since energy has to be added to break a bond (chemical or nuclear), it is convenient to describe the potential energy of bound systems as negative (adding a positive quantity brings the energy of the system to 0). Dreyfus et al. (2014c) have shown that drawing on the *energy as location* metaphor can be productive for developing a conceptual understanding of negative energy, and that experts as well as novices can productively coordinate the *energy as substance* and the *energy as location* metaphors.

While our earlier work shows that students and faculty can productively use both metaphors, in this paper we develop fine-grained cognitive models to demonstrate **how** experts and novices coordinate these two metaphors for energy. One of our central questions in this paper is whether the two metaphors remain separate but coordinated mental spaces, or are better modeled as a single mental space with the concept of energy taking on properties from both metaphors. In exploring this question, we draw on the conceptual blending framework proposed by Fauconnier and Turner (2002), and on gesture analysis (Goldin-Meadow 2003). In addition to showing how ontological metaphors help understand how the concept of energy is built, this paper contributes to the existing literature in two ways: (i) illustrating that conceptual blending can be a useful tool for modeling concepts that cut across disciplinary boundaries, and (ii) illustrating that gestures can be a methodological tool in identifying the character of a blend.

---

[1] These authors contributed equally to the data analysis and production of this paper.

In the following section, we briefly review research on instructional metaphors for energy and situate our research question within this work. Then we review Fauconnier and Turner's framework for conceptual blending and present how it can be applied specifically to generate cognitive models for energy metaphors. We then present our methods, followed by our analysis and results. We end with the implications for research and for instruction.

## Ontological Metaphors for Energy

Within physics education research, energy has long been a topic of interest (Watts, 1983; Lawson & McDermott, 1987; Beynon, 1990; Goldring & Osborne, 1994). Recently, this interest has been renewed, with special emphasis on the metaphors that underlie experts' and novices' reasoning about energy (Amin, 2009; Brewe, 2011; Scherr, Close, McKagan, & Vokos, 2012; Lancor, 2014).

The research on metaphors for energy draws on the theory of conceptual metaphors developed by Lakoff and Johnson (1980/2003, 1999). In Lakoff and Johnson's framework, the way that we think and talk about abstract ideas is organized according to concepts grounded in physical experience. This structuring of one idea or concept in terms of another constitutes a metaphor, and the grounding of chains of metaphors in physical experience is referred to as *embodied cognition*.

Some metaphors are ontological, in the sense that they require mapping ideas across basic categories to which the ideas belong, for example "viewing events, activities, emotions, ideas, etc., as entities and substances" (Lakoff & Johnson, 1980/2003, p. 25). An example (outside of physics) that Lakoff and Johnson present is *the mind is a brittle object*, represented in such sentences as "Her ego is very *fragile*" and "The experience *shattered* him" (Lakoff & Johnson, 1980/2003, p. 28). In a similar vein, *energy as substance* is an ontological metaphor that conceptualizes energy in terms of properties typically attributed to substances/matter, such as thinking of energy as being stored in some object, transferred from one object to another, and conserved (energy cannot be created or destroyed) (Brewe, 2011; Scherr et al., 2012; Lancor, 2014).

A second ontological metaphor for energy conceptualizes it as a vertical location (Amin, 2009; Scherr et al., 2012). In this metaphor, we think of "more energy" as a higher location along a vertical axis (where the vertical axis represents energy values and might not correspond to actual physical positions[2]). In this paper, we focus on how experts and novices coordinate these two metaphors when talking about physical or biological phenomena.

This work on ontological metaphors takes place against the backdrop of an ongoing debate in the literature about the nature of learners' ontologies in physics. A prominent line of research by Chi, Slotta, and their colleagues (Chi & Slotta, 1993; Chi, Slotta, & de Leeuw, 1994; Slotta, Chi, & Joram, 1995; Chi, 2005) argues that physical concepts such as electric current, heat, and light belong to a "scientifically correct" ontological category of emergent processes, that there is a cognitive barrier to flexibly transitioning between ontological categories for a concept, and that robust physics misconceptions about these concepts in novices result from miscategorizing these concepts in the incorrect ontology of substances or direct processes.

---

[2] Typically, the energy axis does not represent location, but in the case of gravitational potential energy in the flat earth approximation, the potential energy is directly proportional to location and the graph of the height looks just like the graph of energy. This example provides a useful pedagogical introduction to this blend.

Recently, this framework has been challenged (Gupta, Hammer, & Redish, 2010; Hammer, Gupta, & Redish, 2011; Brewe, 2011; Scherr, Close, & McKagan, 2012; Jeppsson, Haglund, Amin, & Strömdahl, 2013; Gupta, Elby, & Conlin, 2014; Baily & Finkelstein, Under Review), arguing for a more dynamic view of ontological reasoning. This emerging line of argument demonstrates that experts as well as novices flexibly transition between the substance and process ontologies (including emergent processes) when reasoning about phenomena involving electric current, light, heat, energy, entropy, and quantum mechanics. Physics education researchers have argued for the productivity of conceptualizing non-material entities such as energy, gravity, and heat in terms of properties typically associated with substances (Brookes, 2006; Brewe, 2011; Scherr, Close, McKagan, & Vokos, 2012; Gupta, Elby, & Conlin, 2014). While we do not directly address this debate, our argument here touches on it in two ways: (i) Our observations show how experts and novices can blend ontological metaphors with ease and this lends support to the idea that ontological cognition is flexible; (ii) We add gesture analysis to the methodology of investigating learners' and experts' ontological cognition.

## Conceptual Blending and Metaphors for Energy

In this section we show how metaphors for energy can be conveniently understood within the conceptual blending framework. This forms the background for our subsequent analysis of our video data.

Lakoff's and Johnson's (1980/2003, 1999) conceptual metaphor theory proposes how metaphors arise when one domain (target) is cognitively structured in terms of the elements of another domain (source). The mapping allows aspects of the source domain to be transferred to the target domain. More recently, Fauconnier and Turner (2002) proposed a conceptual blending framework that can also be used to understand the cognitive processes that lie behind metaphors. Their framework shares many aspects with conceptual metaphor theory (Fauconnier & Lakoff, 2013).

In conceptual blending theory, input mental spaces are the cognitive constructs, instead of source and target domains. A blended mental space arises from mappings between entities in the two spaces, and the blended space can (and often does) include entities and relationships from one or both spaces. One key difference, however, is that conceptual metaphor theory mainly deals with entrenched metaphors (where the structuring of one domain in terms of another is relatively stable and long term) (Grady, Oakley, & Coulson, 1999). Blending theory, on the other hand, allows for local mappings between constructs – mappings that might not extend to entire conceptual domains and that might be relatively short-lived. Since we are not dealing with metaphors that are part of our everyday lexicon, or those that we can assume are particularly entrenched (especially in novice reasoning concerning energy), conceptual blending framework allows us greater flexibility in modeling the phenomenon of interest.

In physics education research, conceptual blending has been used to explain analogical scaffolding, the process by which students layer multiple levels of analogies to learn abstract ideas (Podolefsky & Finkelstein, 2007), and to model different ways that students reason about the propagation of wave pulses (Wittmann, 2010).

Understanding how and why conceptual blends occur requires dynamic principles – a model of the causes that drive blends. Fauconnier and Turner (2002) identify critical characteristics of the mental spaces (vital relations) and propose tendencies or forces involving these vital relations that drive blends (constitutive and governing principles). They note that there are many competing tendencies and the dynamical principles of blending theory identify these tendencies rather than predict unique results. Nevertheless, the proposed characteristics and principles (and subsequent elaborations and additions, e.g. (Bache, 2005; Hougaard, 2005)) provide tools for creating new descriptions of cognitive phenomena that can add useful insights.

*Vital relations* are relationships between input spaces that play an important role in conceptual blending. These include such parameters of a situation, such as time, space, cause-effect, part-whole, identity, and representation. One example that Fauconnier and Turner use is that of cold ashes in a fireplace. The way we interpret cold ashes in the fireplace as the remnants of (and resulting from) a past fire represents a conceptual blending: The input space of cold ashes is blended with the input space of fire in the fireplace through compression of various Vital Relations. Both the fire and the ashes occupy the same space but are separated in time (the fire was earlier, the ashes later). From the ashes we can infer there were logs burning in the fire and that it's the logs that became ashes, and that it's the fire that caused that transformation. Thus Fauconnier and Turner illustrate how the two spaces are connected by Vital Relations of Time, Space, Change, and Cause-Effect (Fauconnier and Turner, 2002, p. 96).

Principles identified by Fauconnier and Turner include compressing one vital relation into another, coming up with a story, pattern completion, and unpacking, among others. *Compression* is when we make connections across vital relations between two input spaces; for example, a graph that illustrates different historical recessions (using the same time axis for all of them) is compressing over time. But it is also possible to compress one vital relation into another (e.g., through representing time by space). *Pattern completion* is when we bring in additional elements to the blend by using existing patterns as additional inputs. *Unpacking* is the principle that the blend itself should include all the information necessary to reconstruct the entire network that produced the blend.

For the particular example we consider here, we do not need the full dynamic machinery of blending. The phenomenon we are reporting is well described as using the vital relations of space (the inputs might share locations, or not), part-whole (one input is a part of the other, e.g. when we see a picture of a person's face and identify the person, not just the face, we compress over this vital relation), and representation (one input is a representation of the other, e.g. a picture of an object and the object itself), driven by compression of vital relations, coming up with a story, and unpacking.

Each of the input spaces in a blend has a separate organizing frame. If only one of those frames is projected to organize the blend, then the conceptual integration network is *single-scope*; if the organizing structure of the blended space contains structure that comes from both input spaces, then it is *double-scope* (Fauconnier & Turner, 2002). Single-scope networks include conventional metaphors, in which one input (providing the organizing frame) is the source, and the other input (which is the focus of understanding) is the target. Double-scope networks are more complex, and Fauconnier and Turner have argued that double-scope blending is essential for the development of language and modern human cognition.

## *Conceptual Blending Analysis of Energy*

We describe blended ontologies for energy in terms of two levels of conceptual blending (Figure 1):

First, the *energy as a substance* and *energy as a vertical location* ontological metaphors are described as blended spaces, formed by mapping between the energy input space and the substance and vertical location input spaces. Because these blended spaces take on the organizing structures of the substance and vertical location spaces respectively, they are single-scope blends. If there is a structure to the energy concept in the absence of metaphors (e.g., understanding energy as a purely mathematical construct), this structure is not present in the metaphorical substance and location blends. Second, we hypothesize that the *energy as a substance* and *energy as a vertical location* ontologies become input spaces that form a blended substance-location ontological metaphor. Because this blend incorporates structures of both the substance and location spaces, it is a double-scope blend.

The process of conceptual blending involves selective projection from the input spaces: only some elements from each input space participate in the blend. The full concept of energy is complex and contains many elements; only a subset of these elements is projected into the *energy as a substance* blend, while a different subset is projected into the *energy as a vertical location* blend.

*Describing the Energy as a Substance metaphor as a conceptual blend*

To represent the *energy as a substance* metaphor as a conceptual blend, we first present what the mental space for "substance" or "things" might contain and what the mental space for energy might contain. Of course, these are partial representations of each idea, and the selection of items we represent in these mental spaces are already guided by what we think would be relevant for presenting the metaphor as a conceptual blend. In that sense, this exercise is illustrative rather than explanatory.

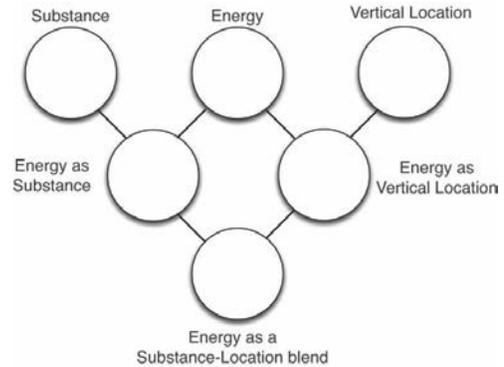

Figure 1: The chain of mental spaces to form the concept of energy

From our everyday experiences, we can conceive of a mental space that incorporates how we think about substances: stuff moves from one place to another, can be stored, can be transformed (e.g., water can freeze into ice), and possesses object permanence (that is, objects cannot simply come into existence from nothing and cannot simply cease to exist).

The energy mental space contains many conceptual elements, which include the ideas that energy can be quantified; the sum total of the energy of a closed system is invariant; energy can be associated with a particular object (or a system of objects)[3]; and there are different types of energy (kinetic, potential, etc.).

As illustrated in Figure 2, mapping elements from the energy space with elements from the stuff space can yield elements of the *energy as a substance* ontological metaphor as identified in Lancor (2014) and elsewhere (Scherr et al. 2012; Brewe 2011).

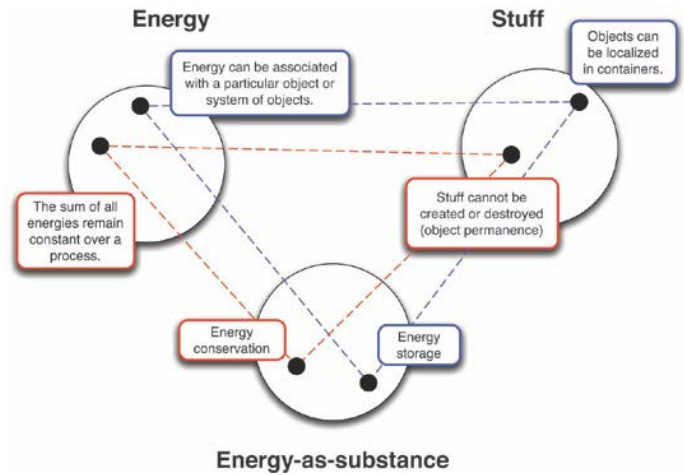

The blended mental space is composed of object (or system of objects) and energy as entities that are now related through the organizational property of "storage," projected from the "Stuff" mental space. Thus in the blended space: *energy is stored* in the associated object (or system of objects). This can also be conceptualized as a compression of the external vital relation between the ideas of association (energy space) and storage (stuff space) into an internal relation within the blended space. Also part of the blend is the mapping of invariance of total energy (of a system) over a process onto object permanence. In the blend, energy (like an object) cannot be created or destroyed (without pre-existing in some form). In doing so, 'energy' and 'stuff' (the contained stuff) are compressed into a uniqueness relation: ***Energy is stuff***.

Figure 2. The energy-as-substance blend.

---

[3] Note that the idea "energy is associated with a particular object" is not a required property of energy, just a conceptual idea that is used in some circumstances. For example, while potential energy may be described as "belonging to a particular object" when one member of the interacting pair is much larger than the other (the potential energy of a thrown ball, or the potential energy of an electron in an atom), in other circumstances energy is associated with multiple objects rather than a single one (the relative kinetic energy of two atoms or the potential energy of an electron-positron pair).

Creating this blend implies that selective properties of the stuff or substances can potentially be projected onto energy. As we "run" the blend, emergent meaning arises from elaboration: a decrease in associated energy of an object is seen as energy *leaving* that object (just as stuff can be taken out of a container, leading to a decrease in the amount of stuff in the container). *Energy conservation* would imply that we now need to account for the energy that leaves an object. The energy could be associated with another object (corresponding to putting stuff into a different container), whose energy thus increases (this is what is seen as energy moving from one object to another - *energy transfer*). Or, the energy is not seen as bound to another particular object (*emission of energy*). Similarly, just like adding stuff to a container, within the blend we can refer to *absorption of energy*. Finally, just as the conservation principle applied to stuff also means that stuff could be transformed from one state to another but not simply cease to exist, conservation of energy can also be understood within the blend through *transformation of energy* from one type to another (for example, the mathematical balance of kinetic and potential energy for an object falling under gravity can be understood in terms of a transformation from potential energy to kinetic energy).

The metaphor of *energy as substance* is so entrenched in our conceptual system, that most of the properties outlined above (emission, absorption, transfer, and transformation) are considered part of how we (experts and novices) talk about energy – properties we can see as having projected back onto the energy space. In what follows, we will use two of these emergent properties of the *energy as substance* space – energy can be absorbed and energy can be emitted – to form a subsequent blend with the *energy as location* space.

*Describing the Energy as a Vertical Location metaphor as a conceptual blend*

The *energy as a vertical location* blend involves a different set of elements from the energy mental space: (i) energy is a state function (i.e., the energy of a system is only a property of the state of the system and is independent of the path that the system took to reach that state), and (ii) the energy of a system is a well-ordered quantity that can be more or less.

The relevant elements of the vertical location mental space are: (i) a vertical 1D coordinate is well-ordered (and a coordinate can be conceptualized as being *higher or lower* than another coordinate), and (ii) a coordinate represents a physical location (objects are "at" locations)

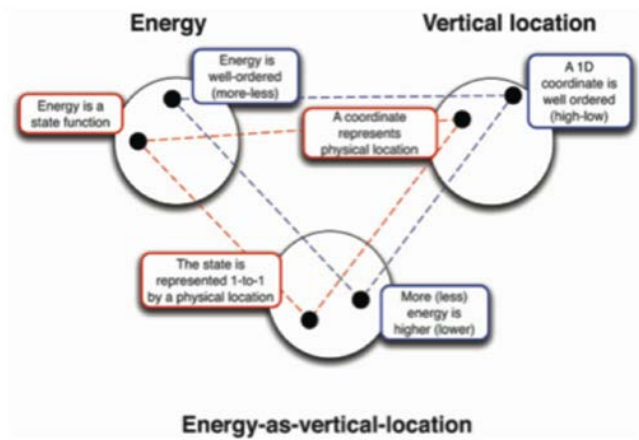

Figure 3. The energy-as-vertical-location blend.

These elements are combined as in Figure 3. In the blended space, energy is mapped onto a vertical location coordinate through a uniqueness compression: energy *is* vertical coordinate. This is represented in the blend through the element "More (less) energy is higher (lower)." Also, in the blended space, the state function of energy represents a physical location, giving meaning to expressions such as "Objects are **at** an energy level." Within the blend, objects with more energy are "higher" than objects with less energy.

As we "run" the blend, new meaning emerges. We know that the negative gradient of the potential energy determines the force vector, which implies that an object (with no initial velocity) will move in the direction of the negative gradient of potential energy. We complete the blend by appending our physical experience with objects in space: that objects tend to fall downwards. Within the blend, new meaning arises, where we can think of the motion of objects within a potential energy gradient as "tending to fall towards lowest potential energy."

*Creating a new mental space through blending the*
*Energy as Substance and Energy as Location metaphors*

We propose that the *energy as substance* and *energy as location* metaphors can themselves be blended to create a new blend that combines properties from both metaphor-spaces. Here, we present what such a blended mental space would look like. Later, we present analysis of verbal data from experts' and novices' reasoning about energy to argue for the psychological existence of the blended mental space.

There are multiple ways that a blend between the substance and location ontologies for energy can manifest itself. We focus on one of these blends, shown in Figure 4, which we observe in our data. In this blend, the idea that objects absorb and release energy (from the *energy as substance* blended space) is mapped onto the idea that more and less energy corresponds to going up and down along a vertical energy dimension (from the *energy as vertical location* blended space). Note that absorption and release of energy were ideas we posited as emerging within the *energy as substance* blended space. These are ideas so entrenched in our everyday conceptualization of energy that we can think of them as reified elements within the *energy as substance* space.

In the new blended space, absorbing energy makes an object go up to a higher energy-location, and releasing energy makes it go down to a lower energy-location. Thus in the blended space, movement along the vertical energy dimension comes to be associated with processes of energy transfer and energy transformation. In this blended space, energy functions simultaneously as both a substance (being moved and transferred) and a vertical location.

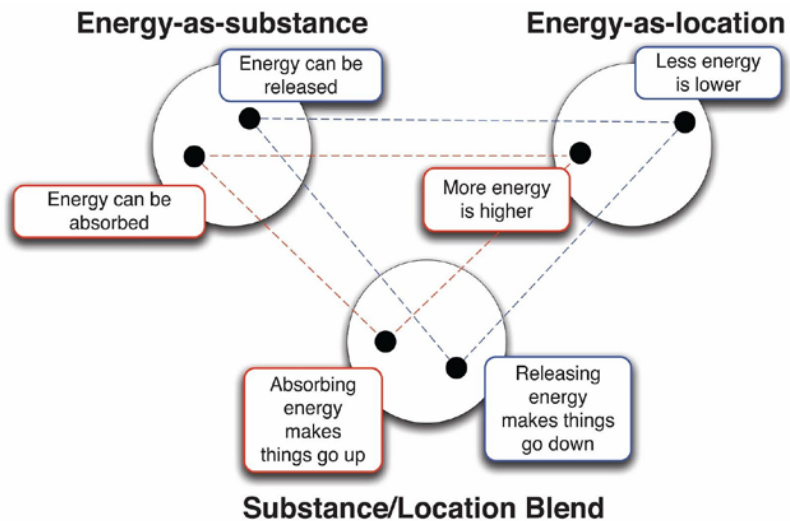

Figure 4. Blending the substance and location blends to describe emission and absorption of energy.

Additionally, new meaning arises through the blending process. Absorption or emission of energy are not just associated with higher or lower energy location, but are seen as processes that cause an object to move up or down the energy axis. This sense of causality emerges within the blend through pattern completion (Fauconnier & Turner, 2003). A more explicit example of this sense of caused motion is apparent in how we explain electron transitions. Figure 5 (Wikipedia, 2014) shows a photon carrying energy out of or into a quantum system by combining the representation of energy in the system as higher and lower levels (location) and the addition or removal of energy as adding or taking away a photon (substance). The electron's movement from one energy level to another is conceptualized as being precipitated by the absorption or emission of the photon.

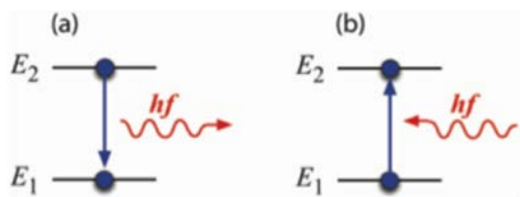

Figure 5. Representation of absorption and emission of energy from the a quantum system. (Wikipedia, 2014).

So far, we have proposed a structure for the conceptual blending of the substance and location ontological metaphors for energy, and now we seek to illustrate what such blended conception might look like when experts or novices explain energy phenomenon. In our analysis of data we draw heavily on gesture analysis, which has previously been a useful tool in science education. In the next section, we briefly discuss the research on gestures and it's use in science education.

## Gestures as tools for cognitive analysis

Gestures are spontaneous hand movements of speakers that are tied to speech and are in the service of communication (Goldin-Meadow, 1999, 2003; Scherr, 2008). Gestures can be classified into multiple categories (McNeill, 1992), including deictic (pointing), iconic (emulating concrete objects), and metaphoric (visually representing abstract concepts).

In science education research, gestures have been analyzed as essential elements of how students communicate and reason about scientific concepts. Gestures play an important role in the development of scientific language and can precede verbal communication (Roth, 2000; Roth & Welzel, 2001; Roth & Lawless, 2002). Gestures can be a source of evidence of the content of students' ideas, particularly for ideas that students are constructing in the moment (Scherr, 2008). Examples include hand motions illustrating the trajectory of a projectile (Scherr, 2008) or the directions of forces and velocities (Chase & Wittmann, 2013).

Gestures can often embody conceptions that are not verbalized in speech. Within conceptual metaphor theory and in conceptual blending theory, metaphors structure not just speech but also the underlying cognition; this implies that non-verbal channels of communicating thinking, such as gestures, could provide additional ways to investigate metaphorical reasoning (Cienki & Müller, 2008). Gestures have been used as part of conceptual blending analysis in previous work, but with a different focus. Edwards (2009), in the context of mathematical reasoning about fractions, used the gestures themselves as one of the input spaces to the blend, and showed how the blended space integrated conceptual and gestural elements. Wittmann (2010) uses gestures as one channel of communication that is analyzed for evidence of conceptual blending. He presents a case study in which students blended one input space of an observed wave pulse on a string with another input space of a thrown ball (leading to the incorrect conclusion that creating a wave on a string with a faster flick of the wrist would cause the wave to move faster). Students' gestures (for example, moving the hand in a motion to produce a wave pulse on a string) served as evidence for this conceptual blending. Our investigation is different from Edwards' (but similar to Wittmann's) in that we are using gestures as evidence for an underlying conceptual blend rather than considering the gestures themselves as an input to the blend.

In the methodology section we describe the types of gestures we specifically draw on to support our analysis of conceptual blending.

# Methodology

## Data Collection and Selection

For this paper, we selected data excerpts from an introductory physics course for undergraduate life sciences majors (Redish et al., 2014). This course is unusual in that it emphasizes chemical energy as a way to build interdisciplinary coherence across physics, biology, and chemistry (Dreyfus et al., 2014a), thereby providing us with opportunities to see how students responded to the concept of energy across disciplinary contexts. We collected video recordings of every class for the first two years. We also conducted semi-structured interviews with students in the course on a variety of topics.

The data set was not collected mainly for the purposes this paper is addressing, but the research questions driving the data were loosely organized around the notion of experts' and novices' ontologies, and hence these data were suitable for mining to address the question we pose here.

The first selection from our data set happened in the course of analyzing the data for purposes outlined elsewhere (Dreyfus et al., 2014c). We viewed video data from classroom and interviews, roughly tagging places where participants tended to reason about energy using substance-based and location-based ideas. This tagging was often done in real time, drawing on verbal phrases such as "you're at this energy" or "put energy into the [object]." For the purpose of the argument presented in this paper, we revisited the tagged episodes and selected out a subset of episodes where both substance- and location-based ideas for energy were in use. These episodes were then analyzed in greater detail as outlined below.

We chose two episodes for analysis:

- One episode from classroom video from Spring 2013. It depicts the physics professor, "Dr. Farnsworth" (pseudonym), reasoning about energy during lecture in service of explaining chemical bond energy to students. The episode caught our attention because of Dr. Farnsworth's use of both substance-based and location-based phrases seamlessly within a short segment of speech.
- A second episode from an interview with a student in the course, "Betsy" (pseudonym), conducted during Spring 2013. Betsy's interview focused on explaining energy transfer and energy transformation in the context of ATP metabolism and other biological processes.

The episodes selected for analysis are intended to be illustrative of what ontological blending looks like. We are not attempting to make empirical claims about the prevalence of blending (relative to using a single ontology, or shifting discretely between two separate ontologies), and therefore we are not concerned with questions of how representative or typical these data are.

## Analysis Methodology

We identify ontological metaphors in our data by looking at two different channels: words and gestures. To analyze the verbal data, we modify slightly the method of Slotta, Chi, and Joram (1995) to fit our particular needs. For gesture analysis, we draw on Goldin-Meadow (2003).

### Predicate Analysis

To analyze the verbal data, we use a version of predicate analysis. Slotta et al. (1995) define predicates as "words, phrases, or ideas whose presence in a spoken explanation is taken to reflect an underlying ontological attribute." Within Chi's and Slotta's framework, attributes belong to ontological categories. The top level categories that Chi and Slotta pose are matter/substance and process, and these are posited to be mutually exclusive (Chi, Slotta, and deLeeuw, 1994; Chi, 2005). The predicates that we use to identify instances of the substance ontology for energy are essentially the same as those used by Slotta et al. (1995). Examples found there include *move* ("energy goes," "energy flows" are example utterances for coding for the *move* attribute in relation to energy), *contain* ("stores energy," "put energy in"), and *quantify* ("a lot of energy," "some of the energy"). A small number of process predicates are found in our data as well, but are not a central part of our analysis. In addition, the other ontological category of relevance to our analysis is location, and predicates that act as markers for location. Location predicates are not included in Slotta et al.'s analysis, so we clarify here what is and is not included in this category.

Our analysis is about ontologies for energy, and so we only code a word or phrase as a location predicate if **energy** is described as a location, and some physical object or system is described as being at (or going to or from) that location. Examples include "here," "go," "up," and "down," when those refer to energy as location. However, "energy goes there" is coded as a **substance** predicate for energy, because the energy is described as being **at** a location (which is an attribute of a substance), not as **being** a location. A location metaphor requires that the location stand for the quantification of the energy, not its spatial location. As a result, the predicates have to be evaluated in context.

*Gesture Analysis*

As with predicates, we posit that gestures can help in providing information on the ontological categories the speaker is drawing on at a given moment. To attend to gestures, we did our video analysis in layers. Analyzing the video jointly, in the first pass, we attended to the content of the speech, trying to make sense of what the speaker was saying. Then we made another careful pass closely attending to the gestures that co-occurred with speech. In this next pass, we would often go through the video frame by frame trying to note the gestures. Here, we posited multiple interpretation of what a particular gesture could be depicting or indexing. In order to understand what meaning participants might be making in a particular moment, we layered the gestures onto the verbal utterance to select which particular interpretations of gestures and speech provided the greatest explanatory power and coherence of meaning.

In our analysis, two kinds of gestures become meaningful: (i) metaphorical and iconic gestures that represent objects and ideas being talked about and/or aspects of the processes being described in speech; (ii) indexical gestures that point to something in the real material setting or in the representational space set up by the speaker at a particular moment. For example, pointing to a location on a physical representation of the vertical energy axis would support the idea that the speaker is thinking of *energy as vertical location*. On the other hand, a gesture tracing the trajectory of energy out of an object (being released into the surrounding air or being transferred to another object) will be taken as indicative of the *energy as substance* metaphor.

Our analysis of gestures is interpretive rather than following a systematic coding scheme. As such, it is difficult for us to make an exhaustive list of what particular gestures are coded for which ontology or metaphor. In lieu of that, we provide enough details in the data analysis so that the readers can draw their own conclusions and evaluate our gesture analysis.

We note that in some of the previous gesture literature (mostly work with children), mismatches of gesture and speech are taken to indicate a destabilization of the child's mental state or a transitional mental state and even a "readiness to learn." A standard example is children performing conservation tasks (water poured from a narrow into a wider glass) using gestures that contradict their words (e.g., showing "wider" when they said "taller"). (Scherr, 2008; Church & Goldin-Meadow, 1986). However, in cases where speech is metaphorical, gesture-speech mismatches might signal something else: that speech is communicating ideas from one mental space (target space, often) while gestures are communicating ideas from a different mental space (source space, often) (Forceville & Urios-Aparisi, 2009). Our interpretation of speech-gesture mismatch here aligns more with the latter interpretation. In our first episode, we are looking at a physics expert using gestures and language to describe an abstract complex concept that is not easily described in physically direct terms. In such situations, experts often use multiple ontologies in their professional interactions (Gupta et al., 2010). Gesture-speech mismatches can indicate that a speaker is thinking about two concepts at once (Chase & Wittmann, 2013), in a way that is more reliable than looking at speech alone, since speech is sequential, while gestures can be another simultaneous channel of communication. We take the gesture-speech mismatches in this case to be evidence of ontological blending, as explained below, and we support our argument with specific examples.

*Inferring Blended Ontologies*

We use predicates and gestures to infer whether participants are drawing on *energy as substance* and *energy as location* metaphors. To infer that participants are drawing on a blended space of these two metaphors we rely on (i) coordination of meaning, (ii) temporal proximity, and (iii) speech-gesture mismatch. When an element from one metaphor space is mapped onto an element from the other metaphor space in the verbal utterance (an identity mapping within the conceptual blending framework) we take that as evidence of a form of compression of an external relation between elements in two different mental spaces into an internal relation within the blended space. Temporal proximity also plays a strong role in this inference. When ideas from two different metaphor spaces are coordinated within the span of a sentence or a phrase, this lends greater evidence that these ideas are not merely coordinated across mental spaces but might be compressed into a single blend. Finally, we note instances in the data where the content of verbal utterance draws from one metaphor space while the co-occurring gestures draw from another metaphor space in a way that meaning-making of the utterance requires us to coordinate the mismatched speech and gesture. Such instances indicate to us that both metaphors are being drawn on simultaneously to express a single integrated idea. Again, such temporal proximity (sometimes less than the span of a second), while not proof that ideas are blended rather than coordinated across the two metaphors, raises the plausibility of the existence of the blended mental space. Within Fauconnier and Turner's framework too, most examples of blended space tacitly rely on such mappings between mental spaces within a short time span.

## Data and Analysis

In this section, we present and analyze transcript segments from the data. The initial predicate analysis has been indicated in the transcripts themselves: substance predicates for energy are <u>underlined</u>, and location predicates for energy are indicated in **bold**. Gestures are also described *(in parentheses and italics)* in the transcript, along with selected still frames shown in the figures.

### *Episode 1: The substance-location blend in expert physics reasoning*

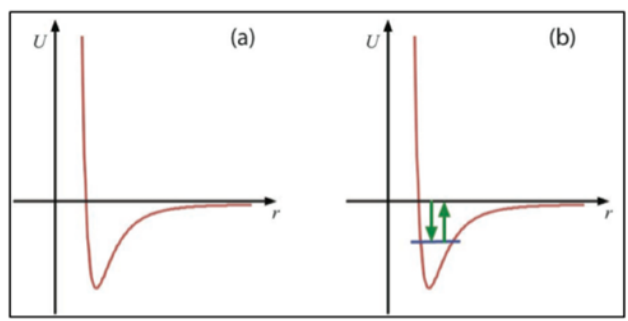

This segment from Dr. Farnsworth comes from the introductory physics for life sciences (IPLS) course. Dr. Farnsworth was going over a quiz in class, and immediately before this segment, he drew a graph of a Lennard-Jones interatomic potential[4] on the chalkboard (see Figure 7 (a)), a representation that had been used frequently in the course for energy associated with chemical bonds. During this segment, Dr. Farnsworth draws additional lines on that graph, shown in Figure 7 (b)..

Figure 7. (a) Lennard-Jones potential and (b) Lennard-Jones potential with the inscriptional marks (horizontal line and vertical arrows) made by Prof. Farnsworth.

Prof. Farnsworth starts with:

> Prof. Farnsworth: If the two atoms are apart *(holds two hands apart, see Figure 8)* and form a bond *(moves hands together)*,

---
[4] The Lennard-Jones potential is a commonly used model of the interaction potential energy between two atoms. (Jones, 1924)

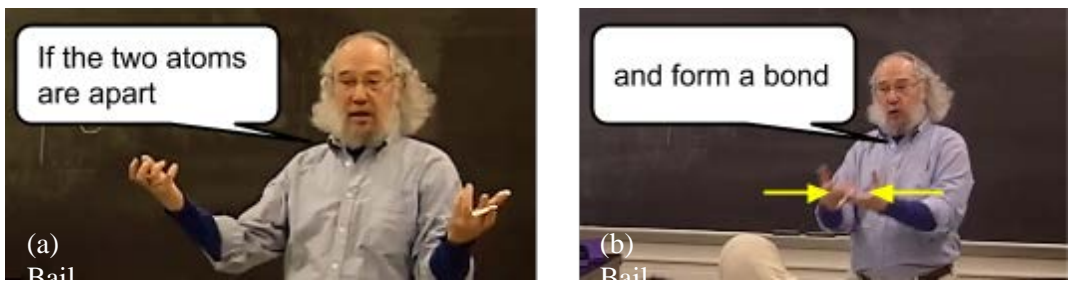

Figure 8. (a) Prof. Farnsworth holds two hands apart, the shape of the hands representing two distinct objects (in this case, atoms). (b) Gesture depicts the atoms coming closer

At the start of the quotation, Prof. Farnsworth held his hands in an iconic gesture as if holding the two atoms in his hands, the distance between his hands depicting the "apart"-ness of the atoms (Figure 8). In the next portion of that sentence, as he said, "and they form a bond," he brought his hands closer. While the gesture depicts the two previously indicated atoms coming closer, the speech only refers to "forming a bond" with no reference to distance. It is only through such speech-gesture coordination that we can make sense of the notion that, as the atoms form a bond, their physical distance is reduced. In the next portion of the utterance, Prof. Farnsworth continues with what this means with respect to energy:

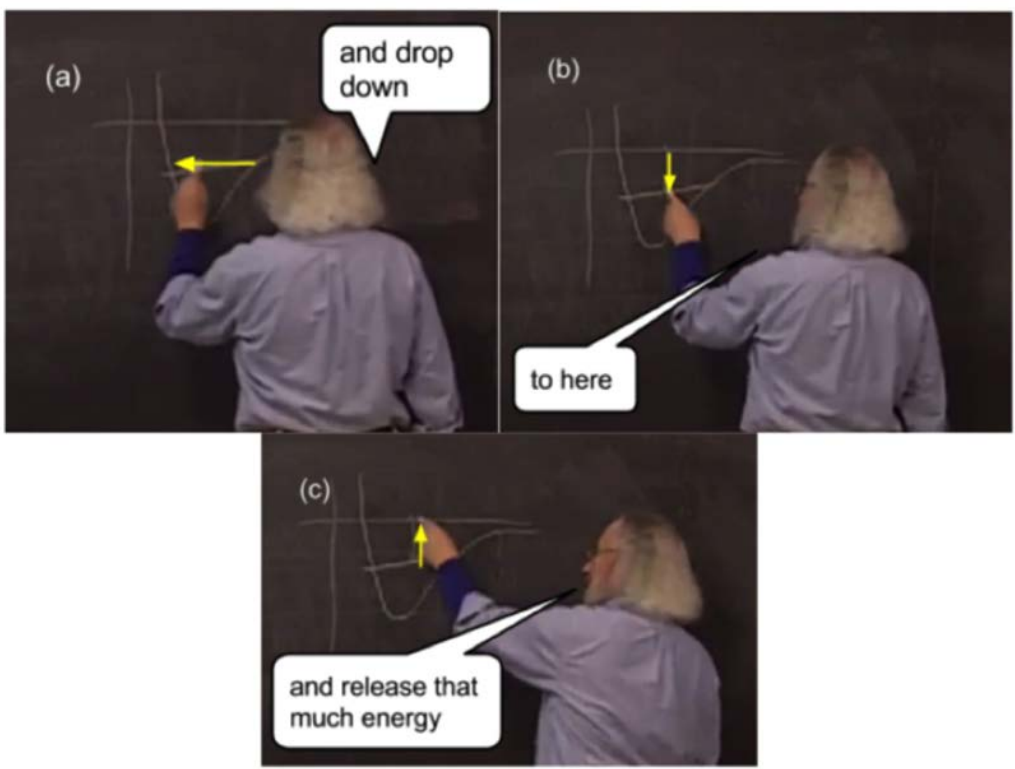

Figure 9. Prof. Farnsworth draws (a) a horizontal line about halfway down the well, (b) a double-tipped vertical arrow from the top of the well to that line, and (c) a vertical arrow from the horizontal line in the middle of the well to the top of the well. (Note that the yellow arrows added to the figure represent his hand motion as he produces the inscriptions on the board.)

Prof. Farnsworth: **they drop down** *(draws a horizontal line about halfway down the well, Figure 9(a))* **to here** *(draws a vertical arrow from the top of the well to that line, see Figure 9(b))* and <u>release that much energy</u> *(see Figure 9(c))*.

As Prof. Farnsworth finished the earlier utterance, he turned to the board. This movement changes the operational space in which he can gesture. Previously, the motion of his hands in the three-dimensional space in front of him was depicting motion of atoms in physical space; the blackboard, on the other hand, has inscribed on it the Lennard-Jones potential as shown in Figure 7(a) in which energy is typically represented on the vertical axis, while physical distance between atoms is represented on the horizontal axis. As we can see from Figure 9(a), the inscription on the board does not mark out the vertical and horizontal axes, but within the context of previous discussions in the course, we can take this as accepted interpretation of those axes.

Simultaneous to saying, "and drop down to here," Prof. Farnsworth draws a horizontal line halfway between the horizontal axis and the bottom point of the Lennard-Jones potential and then draws an arrow pointing downwards from the horizontal axis to the drawn horizontal line. By itself, the utterance ("and drop down to here") could have been a reference to the spatial location of the atoms changing as they form a bond, with atoms dropping in physical space. So, relying only on the verbal utterance, it is unclear whether we can code "drop down" and "to here" as reflecting predicates for the energy-as-vertical-location metaphor. But when we attend to the co-occurring gesture, we can tie the "drop" to the downward-pointing arrow along the vertical energy dimension and "here" as a location on the vertical energy dimension (and not the physical space), leading to the meaning that the atoms, when bonded, are not at a new physical location (in three-dimensional space) but at a new (metaphorical) energy-location. "Drop down" and "to here" can now be coded as predicates for the *energy-as-vertical-location* metaphor. Thus, in this part of the utterance we argue that the fine-grained coordination of speech and gesture are producing the *energy-as-vertical-location* metaphor.

In the next moment, Prof. Farnsworth's speech refers to the amount of energy being released as the atoms form the bond. The word, "release" would refer to the substance attribute "supply" per Slotta et al. (1995), suggesting that the speech in this moment is drawing on the *energy-as-substance* metaphor. However, gestural evidence adds further meaning to this moment: As Prof. Farnsworth utters "released," he draws an arrowhead on the top of the earlier downward pointing arrow, suggesting that the amount of energy released is marked by the difference along the vertical energy dimension. In this coordinated production of speech and gesture, within the span of a second in which it is executed, the speech refers to the *energy as substance* metaphor, while the gestures draw on the *energy as vertical location* metaphor. Additionally, in representing not just energy as vertical coordinate, but also providing a visual sense of the amount of energy released ("that much" represented by the arrow) the inscription on the board becomes a single representational space that binds the two metaphors. In this part of the utterance, we argue that the fine-grained coordination of speech and gesture are producing the substance-location blend for energy.

There is yet another way in which this first sentence of the utterance enacts the substance-location blend. The two phrases, "and drop down to here" and "and release that much energy" are not disjoint phrases that just happen to be in temporal proximity in the utterance, but are conceptually bound so that the full physics implications of the sentence can be understood only by taking the phrases together. The motion of the atoms from one energy-location to another (a change in the energy state function of the system) is causally linked with the release of energy. This kind of causal elaboration is an emergent property of the substance-location blend as described before. This causal link becomes more explicit in the following segment:

Prof. Farnsworth: And because **that's where they are** *(holds hands with the palms down, imitating the shape of the potential well)*, **at that negative energy**, that's equal to the energy you have to put in to get them back apart *(gestures getting the atoms apart)*. So it's just about *(puts one hand above the other, and moves hands up and down, out of phase)* **where you're going** *(turns to chalkboard)*, that when you form a bond *(puts hand horizontally at top of well)*, you're **dropping down** *(moves hand downward to horizontal line in middle of well)*, and if you **come in at this energy** *(redraws horizontal line at top of well)* you gotta get rid of this much *(draws another vertical arrow from* *the top of the well to the horizontal line in the middle).* But if you're **down** *(holds hand horizontally at top of well)* **here** *(moves hand to horizontal line in middle of well)* and want to **get back up to here** *(moves hand back to top of well)*, you gotta put in this much *(draws vertical arrow from horizontal line in middle of well to top of well, see Figure 10).*

Note that the phrasing "if you… get back up here, you gotta put in that much [energy]" indicates a causal dependence between energy transfer and motion of atoms along the vertical axis of energy.

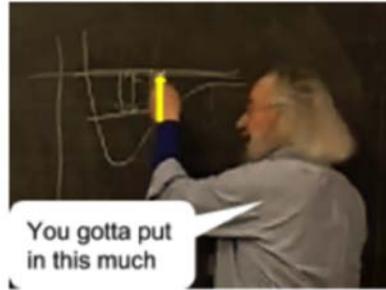

In this segment, the pattern of verbal utterance and gestures that we detailed above is sustained: (i) substance and location predicates are interlaced within the same sentence, (ii) releasing (or putting in) energy is mapped onto moving down (or up) in energy-location, and (iii) the physical inscription of the Lennard-Jones potential is maintained as a space that represents energy-as-vertical location with arrows and distances along that axis representing amounts of energy released or put in. This indicates that Prof. Farnsworth is not just switching back and forth between the two metaphors, but that ideas from the two different mental spaces are blended into a single mental space. In this blended space, adding (or removing) energy to the system of two atoms doesn't just increase their energy; it moves them to a higher (or lower) energy-location. The contractions from individual mental spaces to the blended space are reflected in the speech-gesture "mismatch," where the predicate from one ontological category is accompanied by gestures from another.

Figure 10. Professor Farnsworth draws upward arrow to indicate the amount of energy needed to break the bond.

## *Episode 2: The substance-location blend in novice physics reasoning*

Betsy was an undergraduate pre-medical student enrolled in the introductory physics for life sciences course taught by Prof. Farnsworth. This interview was the second of two interviews with her that focused on reasoning about energy. The interviewer drew two Lennard-Jones potential graphs on the whiteboard (see Figure 11), told Betsy that they represent two different systems, and asked Betsy which graph represented more energy. The question was intentionally underspecified: on the one hand, the minimum of the graph on the right (representing a bound state) is at a higher energy than the minimum of the graph on the left. On the other hand, the difference in energy between the bound state and the unbound state (shown as zero on the graph) is greater for the graph on the left. Betsy saw through the trick question and explained why either graph could be the answer, depending on the interpretation of the question.

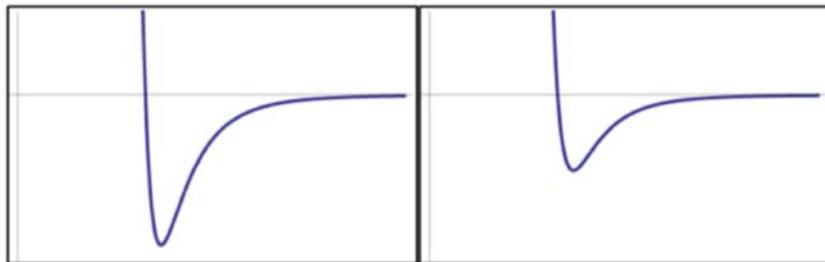

Figure 11. Lennard-Jones potential graphs for two interacting atomic systems, as drawn in the Betsy interview. The vertical axis represents energy, and the horizontal axis represents the distance between atoms.

The interviewer next followed up on an earlier conversation about ATP hydrolysis, and asked:

**Interviewer:** So between these two graphs, if one of these represented ATP and one of them represented ADP which one would you say was which?

**Betsy:** *(immediately writes ATP on the right and "ADP+P in water" on the left)*

**Interviewer:** Okay, why?

**Betsy:** Because in biology they always assume that it's in water 'cause the whole system is mostly made up of water. So if I put these two graphs together *(draws new graph with two valleys, see Fig. 12)*, so this is ATP *(labels shallower well as "ATP")* and it takes a little bit of energy to put in to get ADP *(traces up along graph from ATP well and then back down into deeper well, and labels deeper well as "ADP", see Fig. 13)*, but ADP is much more stable than this *(points to ATP)*, and this is because the phosphate reacts with the water and forms a really stable. So it's **in a well** but it **falls into a deeper well** once the bond breaks. I'm pretty sure.

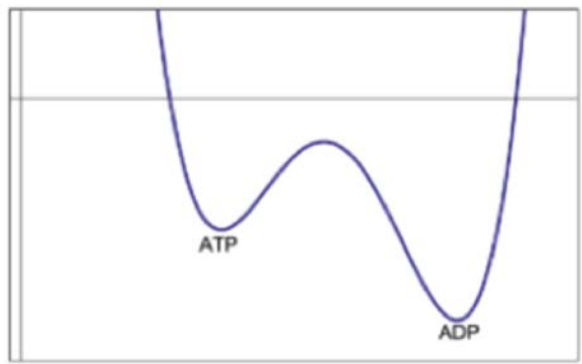 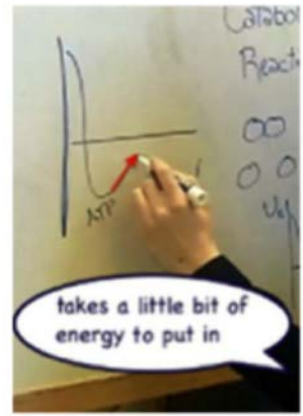

Figure 12. The energy diagram that Betsy draws, representing ATP and ADP on the same graph

Figure 13. Betsy traces up along graph from ATP well and then back down into deeper well (labeled as ADP)

Here, Betsy's co-occurring gestures and speech draw on resources from different ontological metaphors. As she says, "it takes a little bit of energy to put in" (reflecting the quantity and absorb attributes, per Slotta et al. (1995)), she traces a path upward along the graph to indicate what happens to the system when energy is "put in." The vertical location on the graph represents energy. So we interpret Betsy's actions here to mean that, as energy is put in, the system moves up in energy-location. This location ontology is reinforced with Betsy's subsequent utterance that the system is "in a well" and "falls into a deeper well" (the well is not a spatial location, but a location along the energy dimension). The coordinated speech and gestures drawing simultaneously on "energy can be absorbed or put into an object" and "objects are at energy locations" reflect the blended element "absorbing energy makes things go up."

A few turns later in the conversation, Betsy says, "They call ATP a high-energy bond," and the interviewer follows up on this:

**Interviewer:** Yeah, why do they—what does that mean when they call it a high energy bond?

**Betsy:** Yeah, it doesn't actually have—Professor [Farnsworth] was talking about that, it doesn't actually hold energy *(gestures as if "holding" something, see Figure 14(a))*, like it's not—like, the bond itself doesn't have a lot of energy, but it's the fact of breaking it and forming this *(points to ADP)* is even more—is **even lower energy** *(gestures up and down, see Figure 14(b))*. So you can get—the difference between **here** and **here** *(labels vertical difference between two wells, see Figure 14(c))*, this is the energy that's given off to drive the rest of the system. Or drive the rest of the reaction.

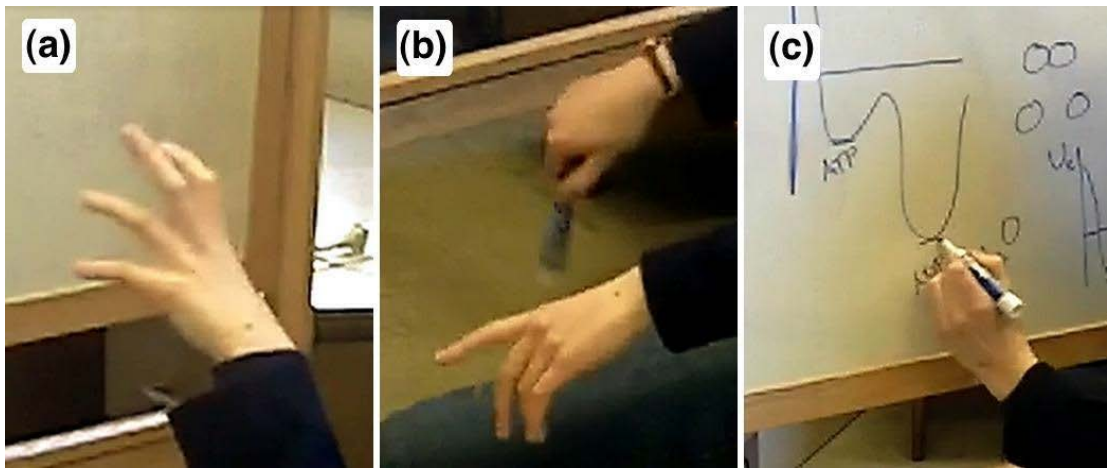

Figure 14. (a) Betsy gesturing for "doesn't actually hold energy"; (b) Betsy gesturing for "is even lower energy; (c) Betsy's diagram for "between here and here".

Here, again, we see Betsy interleaving substance and location ideas (reflected in her verbal utterances and gestures). "Hold energy," "have a lot of energy," and "energy that's given off" indicate contain, quantify, and supply substance-attributes (per Slotta et al. (1995)). The utterance of "even lower energy" is coordinated with gestures to indicate lower energy-location. And the difference in positions of the energy locations ("the difference between here and here") is referred to as the amount of energy that is released. We take this close integration of the two metaphors as evidence of blending the two mental spaces to produce the idea (in blended space) that "releasing energy makes things go down."

This blending enables representing energy conservation in the location space. While conservation was already an affordance of the substance space (putting in stuff from elsewhere increases the total amount of stuff inside, and losing stuff to the outside decreases the total amount of stuff on the inside), conservation is not represented in the location space on its own (going to a higher or lower location is not obviously associated with changes in the surrounding world). However, in the blended space, changes in location ("the difference between here and here") are associated with the transfer of a substance ("the energy that's given off") and therefore with interactions with the surroundings.

## Conclusions and Implications

We present illustrative cases on how a physics professor and an undergraduate student in an interdisciplinary introductory physics course for life sciences majors draw on two different metaphors for energy. We argue that for both of these subjects, the two metaphors are integrated into a single blended mental space as evidence through our analysis of their overlapping speech and gestures.

Our results offer further support for the value of viewing ontologies in physics not simply as fixed, but as a dynamic tool for building powerful conceptual models of abstract and difficult concepts. (Gupta, Hammer, & Redish, 2010; Hammer, Gupta, & Redish, 2011; Gupta, Elby, & Conlin, 2014). However, our analysis here does not directly address the argument for ontological commitments leading to misconceptions, as presented by Chi and Slotta (Chi, 2005; Slotta & Chi, 2006), since their argument pertained to the categories of substance and emergent processes, rather than that of location.

Through this paper we also contribute to the growing trend of multi-modal analysis (Stivers & Sidnell, 2005), attending to not just verbal utterances of learners but also other embodied modalities, to create models of learners' cognition. We illustrate one possible way in which speech and gesture can be co-analyzed to make sense of learners' and experts' cognition. In particular, we show how Fauconnier and Turner's conceptual blending theory in conjunction with gestural analysis can provide useful machinery for understanding how novices and experts blend multiple metaphors when reasoning about science concepts such as energy.

As has been well documented, expert knowledge (Chi, Feltovich, & Glaser, 1981; Reif, 1995; Machamer, Darden, & Carver, 2000) is not just about learning sets of independent facts; it is about organizing and coordinating facts and principles, seeking local and global coherence, and being able to generate sensible stories about complex phenomena. If we are to be able to understand students' progress along this dimension, we need analytic tools that can help us both understand expert knowledge structures and make sense of how students acquire and create such structures. While a knowledge-in-pieces theoretical frame (diSessa, 1993; Smith, diSessa, & Roschelle, 1994; Hammer, 1996) can help us make sense of the resources students can activate when faced with a learning task, we need more analytic tools to see how these resources interact (Gupta & Elby, 2011) and generate new knowledge structures. We believe that cognitive blending offers one such tool. (See also Wittmann, 2010.) The sort of analysis we have done here, using the machinery of conceptual blending to model the dynamics of a learner's cognition can potentially help us give us insight into how students learn to build new knowledge.

In particular, many science concepts such as energy are understood primarily through metaphors, often through multiple metaphors. Expertise in physics is not constituted in learning a single canonical way of reasoning about a concept or ontologically categorizing that concept; rather, it is constituted in coordinating multiple metaphors and ontological categories to flexibly understand and apply that concept in different contexts. We expect that the kind of analysis done here can not only provide a better understanding of what students are doing in their path towards developing expertise, but also offer guidance in how to create learning environments that facilitate that development. Our previous work provides an example of this. The context was the NEXUS/Physics course described above and the instructional goal was the extension of the traditional treatment of energy in introductory physics to include a discussion of chemical energy and bonding (Dreyfus et al. 2014a). Our ontological research with this population (Dreyfus et al. 2014c) and our instructional experience shows that many students apply the "energy as substance" metaphor and have difficulty making sense of negative potential energy in general and binding energy in particular. This led us to emphasize the fact that the zero of potential energy is arbitrary and in some cases, convenience leads to a choice that makes potential energy and even total energy negative. These lessons provide a scaffold for students to build the metaphor of "energy as location". This is done at the beginning of the discussion of energy in traditional contexts so it is not a new idea when the topic comes up in the context of chemical bonding. A critical element of this instruction is the use of visual representations that allow a geometric interpretation of the negatives (potential energy graphs with negative values and energy bar charts with bars that can be positive or negative).

For the current results, we plan to extend our previously developed examples and add new ones to focus on adding and removing energy. This will mean bringing in visual representations that blend the metaphors (such as figures 5 and 7). The exposure to the blended energy metaphor will prepare students for later work on spectroscopy in chemistry where such diagrams are used extensively and quickly become quite complex.

This illustrates how a careful theoretical analysis of cognitive models can work with instructional development to better integrate and reconcile distinct mental models, as often occurs, for example, in articulating instruction among different disciplines (Dreyfus et al. 2014b).

## Acknowledgements

This work was supported by the National Science Foundation under the Graduate Research Fellowship (DGE 07-50616), NSF grant DUE 11-22818, and the Howard Hughes Medical Institute NEXUS grant. The authors thank the rest of the NEXUS/Physics research team (Ben Geller, Julia Gouvea, Vashti Sawtelle, and Chandra Turpen), the UMD Physics Education Research Group, Michael Wittmann, and two anonymous reviewers for discussions and feedback.